\documentclass[12pt]{article}
\usepackage{epsf}

\catcode`@=11
\def\un#1{\relax\ifmmode\@@underline#1\else
        $\@@underline{\hbox{#1}}$\relax\fi}
\catcode`@=12

\def\a{\alpha}
\def\b{\beta}

\def\d{\delta}
\def\e{\epsilon}
\def\f{\phi}

\def\h{\eta}

\def\j{\psi}
\def\k{\kappa}
\def\l{\lambda}
\def\m{\mu}

\def\o{\omega}
\def\p{\pi}
\def\q{\theta}

\def\D{\Delta}

\def\O{\Omega}

\def\S{\Sigma}

\def\bo{{\raise-.5ex\hbox{\large$\Box$}}}               
\def\pa{\partial}                                       
\def\pr{\prod}                                          
\def\TH{{\raise.2ex\hbox{$\displaystyle \bigodot$}\mskip-4.7mu \llap H \;}}
\def\face{{\raise.2ex\hbox{$\displaystyle \bigodot$}\mskip-2.2mu \llap {$\ddot
        \smile$}}}                                      

\def\sp#1{{}^{#1}}                              
\def\ket#1{\left| #1\right\rangle}              
\def\VEV#1{\left\langle #1\right\rangle}        
\def\leftrightarrowfill{$\mathsurround=0pt \mathord\leftarrow \mkern-6mu
        \cleaders\hbox{$\mkern-2mu \mathord- \mkern-2mu$}\hfill
        \mkern-6mu \mathord\rightarrow$}
\def\dvec#1{\vbox{\ialign{##\crcr
        \leftrightarrowfill\crcr\noalign{\kern-1pt\nointerlineskip}
        $\hfil\displaystyle{#1}\hfil$\crcr}}}           
\def\dt#1{{\buildrel {\hbox{\LARGE .}} \over {#1}}}     

\def\frac#1#2{{\textstyle{#1\over\vphantom2\smash{\raise.20ex
        \hbox{$\scriptstyle{#2}$}}}}}                   
\def\ha{\frac12}                                        
\def\sfrac#1#2{{\vphantom1\smash{\lower.5ex\hbox{\small$#1$}}\over
        \vphantom1\smash{\raise.4ex\hbox{\small$#2$}}}} 
\def\bfrac#1#2{{\vphantom1\smash{\lower.5ex\hbox{$#1$}}\over
        \vphantom1\smash{\raise.3ex\hbox{$#2$}}}}       
\def\afrac#1#2{{\vphantom1\smash{\lower.5ex\hbox{$#1$}}\over#2}}    

\def\[{\lfloor{\hskip 0.35pt}\!\!\!\lceil}
\def\]{\rfloor{\hskip 0.35pt}\!\!\!\rceil}

\def\un{\underline}
\def\fracmm#1#2{{{#1}\over{#2}}}

\def\low#1{{\raise -3pt\hbox{${\hskip 0.75pt}\!_{#1}$}}}

\def\Dot#1{\buildrel{_{_{\hskip 0.01in}\bullet}}\over{#1}}
\def\dt#1{\Dot{#1}}

\newskip\humongous \humongous=0pt plus 1000pt minus 1000pt
\def\caja{\mathsurround=0pt}
\def\eqalign#1{\,\vcenter{\openup2\jot \caja
        \ialign{\strut \hfil$\displaystyle{##}$&$
        \displaystyle{{}##}$\hfil\crcr#1\crcr}}\,}
\newif\ifdtup

\def\ref#1{$\sp{#1)}$}

\def\pl#1#2#3{Phys.~Lett.~{\bf {#1}B} (19{#2}) #3}
\def\np#1#2#3{Nucl.~Phys.~{\bf B{#1}} (19{#2}) #3}
\def\prl#1#2#3{Phys.~Rev.~Lett.~{\bf #1} (19{#2}) #3}
\def\pr#1#2#3{Phys.~Rev.~{\bf D{#1}} (19{#2}) #3}

\def\jmp#1#2#3{J.~Math.~Phys.~{\bf {#1}} (19{#2}) #3}

\def\mpl#1#2#3{Mod.~Phys.~Lett.~{\bf A{#1}} (19{#2}) #3}

\parskip=\medskipamount                 
\thicklines                         

\def\pco{P\!C\!O}
\def\sfo{S\!F\!O}
\def\ico{I\!C\!O}
\def\cross{\nearrow\kern-1em\searrow}
\def\contract#1#2{\vtop{\ialign{##\crcr
   $\hfil\displaystyle{#2}\hfil$\cr\noalign{\kern-6pt} \kern3pt\vrule
   height#1pt\hrulefill \vrule height#1pt\kern3pt\crcr}}}

\begin{document}


\begin{titlepage}

\noindent
DESY 96--177 \\
ITP--UH--19/96 \\
hep-th/9608196 \hfill August 1996\\

\vskip 0.6cm

\begin{center}

{\Large\bf Restoring Reality for the Self-Dual N=2 String~$^*$}\\

\vskip 1.5cm

{\large Jan Bischoff \ and \ Olaf Lechtenfeld}

\vskip 0.6cm

{\it Institut f\"ur Theoretische Physik, Universit\"at Hannover}\\
{\it Appelstra\ss{}e 2, 30167 Hannover, Germany}\\
{http://www.itp.uni-hannover.de/\~{}lechtenf/}\\

\vskip 2cm
\textwidth 6truein
{\bf Abstract}
\end{center}

\begin{quote}
\hspace{\parindent}
{}\ \ \ 
It is known that the critical $N{=}(2,2)$ string describes 
$2{+}2$ dimensional self-dual gravity in a non-covariant form,
since it requires the choice of a complex structure in the target,
which leaves only $U(1,1)$ Lorentz symmetry.
We briefly review picture-changing and spectral flow and 
show that the world-sheet Maxwell instantons individually 
break the Lorentz group further to $SU(1,1)$. 
However, their contributions conspire to restore full
$SO(2,2)$ global symmetry if dilaton and axion fields are
assembled in a null anti-self-dual two-form, 
denying them the status of Lorentz scalars.
We present the fully $SO(2,2)$ invariant tree-level three-point
amplitude and the corresponding extension of the Plebanski action
for self-dual gravity.
\end{quote}

\vfill

\textwidth 6.5truein
\hrule width 5.cm

{\small
\noindent ${}^*$
Supported in part by the `Deutsche Forschungsgemeinschaft'
}

\eject
\end{titlepage}

\newpage
\hfuzz=10pt

\section{Introduction and Results}

Closed strings with $(2,2)$ world-sheet supersymmetry are 
perhaps the only exactly solvable four-dimensional closed string theories,
providing a consistent quantum theory of self-dual 4d gravity~\cite{ma}.
This hope derives from the observation 
that the spacetime background of signature~$2{+}2$ leaves no room
for perturbative transverse string excitations 
and only a massless spectrum remains. 
Consistency of the absence of massive poles with duality then
requires the perturbative vanishing of all string scattering amplitudes 
beyond the three-point function.
In particular, the crucial vanishing of the four-point function
is tied to the peculiar kinematics in $2{+}2$~dimensions.
This is as simple as a string theory can get.
Still, one has to deal with the perturbative string expansion which,
due to the presence of the graviphoton
of the $N{=}2$ world-sheet supergravity,
sums over world-sheet genera as well as Maxwell instanton numbers.

Five years ago, Ooguri and Vafa~\cite{ov1} showed that 
the single massless physical field degree of freedom~$\f$ 
present in the closed $N{=}2$~string 
parametrizes the K\"ahler potential of self-dual gravity
in $2{+}2$~dimensions.
After computing tree-level amplitudes with up to four legs 
they indeed identified the Plebanski action~\cite{ple} of self-dual gravity
as the (tree-level) effective spacetime action of the closed $N{=}2$~string.
Since self-dual structures in four dimensions are believed to
unify all integrable systems in two and three dimensions,
$N{=}2$~strings should also teach us about the quantization
of integrable models.

Two years ago, Berkovits, Ooguri and Vafa~\cite{bv2,ov2} 
reformulated type~II $N{=}2$~strings in terms of $N{=}4$ topological strings. 
They showed that a rotation of the spacetime complex structure
(i.e. the choice of $R^{2,2}\to C^{1,1}$)
mixes the amplitudes from the various Maxwell instanton numbers
occurring at a given genus.
Since such a (so-called flavor) rotation is nothing but a transformation
under the $SU(1,1)_f$ factor of the would-be Lorentz group 
$SO(2,2)=[SU(1,1)_c\otimes SU(1,1)_f]/Z_2$,
fixed-instanton-number amplitudes are in general only invariant
under (so-called color) $SU(1,1)_c$.
Only for zero instanton number do the amplitudes share the full global
$[SU(1,1)_c\otimes U(1)_f]/Z_2\otimes Z'_2$
symmetry of the Brink-Schwarz action~\cite{bsa}.
This fact has been supported by explicit computations~\cite{lp,kl,buck}
and has led to some controversy.
In particular, it raised the question whether it is possible 
to restore the global $SO(2,2)$~invariance, which had to be broken
by choosing a complex structure in order to write down the
Brink-Schwarz action in the first place.

In this letter we point out a way to restore $SO(2,2)$ invariance.
We make use of the freedom in the definition of any fermionic string theory
which arises when putting together
left- and right-moving monodromies for world-sheet fermions.
It will also be necessary to invoke the dynamical nature of the
string couplings $\k$ and~$\l$ 
which weigh the different world-sheet topologies.

The genus~$g$ of the world-sheet~$\S$ and the instanton number~$c$
of the principal $U(1)$ bundle over~$\S$ are given by
$$
\frac{1}{2\p}\int_\S R \ =\ 2-2g \qquad,\qquad
\frac{1}{2\p}\int_\S F \ =\ c \quad,
\eqno(1) $$
where $R$ and $F$ are the curvature two-forms of the spin and Maxwell
connections of $N{=}2$ world-sheet supergravity, respectively.
For each topology, labeled by~$(g,c)$, 
there are metric, fermionic, and Maxwell moduli spaces
to be integrated over~\cite{kl}.
The integral over the Maxwell moduli of flat $U(1)$ connections
combines with the spin structure sum to a {\it continuous\/} sum
over world-sheet fermionic monodromies.
It turns out that nothing depends on the NS or R
(or interpolating) assignments for external states,
a manifestation of the {\it spectral flow\/} endomorphism
of the $N{=}2$ superconformal constraint algebra.

Our definition of the $N{=}2$ string declares 
left and right monodromies as independent and, consequently, 
sums over left and right spin structures separately.
The same prescription turns the $N{=}1$ closed fermionic string 
into the type II superstring by way of the GSO projection.
For the $N{=}2$ string, 
it amounts to having independent right and left spectral flow symmetries
and is compatible with modular invariance.
External state vertex operators can be twisted by a pair of
spectral-flow operators
$$
V(k)\ \longrightarrow\ V^{(\q_L,\q_R)}(k)\ :=\
\sfo_L(\q_L)\,\sfo_R(\q_R)\,V(k) \quad,
\eqno(2) $$
and it is easy to see that correlators of vertex operators are unchanged
as long as their total twists (left and right) vanish~\cite{kl}.
Nonzero integral total twists $(\q_L,\q_R)=(c_L,c_R)$, however,
shift the instanton number, i.e. they amount to a topology change~\cite{buck}!
Even though geometrically not obvious, 
we are led to sum over a {\it pair\/} of instanton numbers 
in the expression for the full $n$-point scattering amplitude,
$$
A(k_1,\ldots,k_n)\ =\
\sum_{g\in Z_+}\k^{2g-2+n}\sum_{c_L\in Z}\sum_{c_R\in Z} 
\ell_{c_L,c_R}\,\l^{-c_L-c_R}\;A^g_{c_L,c_R}(k_1,\ldots,k_n)\quad,
\eqno(3) $$
with a priori unknown integral multiplicities $\ell_{c_L,c_R}$, 
to be fixed later.~\footnote{
We might absorb those into $A^g_{c_L,c_R}$ but the
latter are naturally related by spectral flow.}
Because $(A^g_{c_L,c_R})^*=A^g_{-c_L,-c_R}$,
reality demands that $\k$ is real and $\l$ is a phase.
It should be noted that fermion zero modes make $A^g_{c_L,c_R}$ vanish
for any $|c|>2g{-}2{+}n$, restricting the instanton sum in eq.~(3)
to $|c_L| \le 2g{-}2{+}n \ge |c_R|$.

Since flavor rotations of $A$ mingle the $A^g_{c_L,c_R}$ for $g$~fixed,
$SO(2,2)$~invariance can be achieved only if the string couplings 
$\k$ and~$\l$ are allowed to vary in compensation!
This seems strange at first since, 
in a background more general than $C^{1,1}$, we should identify
$$
\k\ =\ e^{\VEV{d}} \qquad,\qquad
\l\ =\ e^{i\VEV{a}} \quad,
\eqno(4) $$
where $d$ and~$a$ stand for the spacetime dilaton and axion fields,
which couple to $R$ and~$F$, respectively.
One is used to view these fields as real spacetime Lorentz scalars.
Here, we propose unusual Lorentz properties for the dilaton and axion fields!
Using the isomorphism $SU(1,1)/Z_2=SO^+(2,1)$ for the flavor subgroup, 
we find that the triple~$w$ given by
$$
\left(\begin{array}{c} w^1 \\ w^2 \\ w^3 \end{array}\right)\ :=\ \sqrt{\k} 
\left(\begin{array}{c}(\l{+}\l^{-1})/2\\(\l{-}\l^{-1})/2i\\1\end{array}\right)
\ =\ \exp\{\VEV{d}/2\}\,
\left(\begin{array}{c} \cos\VEV{a} \\ \sin\VEV{a} \\ 1 \end{array}\right)
\eqno(5) $$
must transform as a (massless) vector in $2{+}1$ Minkowski space.
Other options encode the dilaton and axion degrees of freedom
in a Majorana-Weyl spinor~$v$ of $SO(2,2)$ 
or a null anti-self-dual two-form~$\O^-$
in the full $2{+}2$ dimensional background.
We suggest that the lack of a covariant action for self-dual gravity 
may be overcome when such a two-form is employed with the metric.
Alternatively, the presence of an $SO(2,2)$ spinor hints at the
possibility of spacetime supersymmetry (see, however, refs.~\cite{sie,klp}).

In the following, we shall show how the Lorentz transformations
of $\k$ and~$\l$ arise naturally already on the level of the vertex operators, 
i.e. from BRST cohomology.
After a discussion of physical states in the different pictures of the 
$N{=}2$ string and their relation by picture-changing and spectral flow, 
we detail the behavior of the theory under $SU(1,1)_f$ rotations,
which change the complex structure and complete the would-be Lorentz group.
We demonstrate how the instanton sum can restore $SO(2,2)$ invariance
of the scattering amplitudes.
Finally, the tree-level three-point function is worked out completely,
and the corresponding spacetime action is constructed, providing a
stringy extension of self-dual gravity.

\section{ Physical States, Pictures and Spectral Flow }

The physical states or vertex operators for the type~II $N{=}2$~string
are obtained as elements of the relative BRST cohomology, meaning
that one imposes as subsidiary conditions the vanishing of the
reparametrization and Maxwell anti-ghost zero modes on physical states.
This cohomology factorizes into the relative BRST cohomologies 
for the two chiral halves of the string. 
Let us consider the left-moving cohomology, dropping the $L$~subscript. 
It is graded by
\begin{itemize}
\item conformal dimension $h$ as eigenvalue of $L_0^{\rm tot}$;
      physical states must have $h=0$.
\item local Maxwell charge $e$ as eigenvalue of $J_0^{\rm tot}$;
      physical states must have $e=0$.
\item picture numbers $(\p^+,\p^-)$, with $\p^+{+}\p^-\in Z$ and
      $\p^+{-}\p^-\in R$, 
      labeling inequivalent superconformal ghost vacua. 
\item total ghost number $u\in Z$ 
      as eigenvalue of the total ghost charge~$U$;
      physical states are presumed to occur only for
      $\tilde u\equiv u{-}\p^+{-}\p^{-}=1$.
\item color quantum numbers $(j,m)$ labeling $SU(1,1)_c$ behavior;
      physical states are believed to be singlets.
\item global $U(1)_f$ charge~$q\in R$; it will play a key role.
\item complex center-of-mass momentum $k^\m$ 
      as eigenvalue of $P^\m=\frac1{2\p}\oint\partial X^\m$;
      physical states are massless, i.e. $k^\m k^*_\m=0$.
\end{itemize}
The relative BRST cohomology has been worked out 
for the ``canonical'' picture of $(\p^+,\p^-)=({-}1,{-}1)$, 
finding indeed a single massless physical state 
$\ket{\f}=V_{\rm can}(k)\ket{0}$ with 
$u={-}1$, $(j,m)=(0,0)$, and $q=0$~\cite{bien,bkl}. 

It is very helpful that there exist BRST-invariant operators
which carry non-zero picture numbers and thus may relate the
cohomologies in the various pictures. One has picture-changing~\cite{fms,bkl}
and spectral-flow~\cite{bv2,kl} operators, with charges 
\vglue.1in \noindent
\hfill
\begin{tabular}{|l|c c c c c c|} \hline
operator & $\p^+$ & $\p^-$ & $u$ & $v$ & $(j,m)$ & $q$ \\ \hline
$\pco^+$ & 1 & 0 & 1 & 0 & (0,0) & 0 \\
$\pco^-$ & 0 & 1 & 1 & 0 & (0,0) & 0 \\
$\sfo(\q)$ & ${+}\q$ & ${-}\q$ & 0 & 0 & (0,0) & ${+}\q$ \\ \hline
\end{tabular}
\hfill (6) 
\vglue.1in \noindent
and $h=e=0=k^\m k^*_\m$ for everybody. 
It is important to note that $\pco^+$, $\pco^-$, and $\sfo(\q)$
commute with one another, up to BRST-exact terms.

Since $\sfo(\q,z)=\exp\left[2\,\q\int_{z_0}^z J^{\rm tot}\right]$,
spectral flow is additive and therefore invertible via $\q\to{-}\q$.
Hence, it provides a one-to-one map between cohomology classes for
the same value of $\p\equiv\p^+{+}\p^-$.
In particular, in all $\p{=}{-}2$ sectors
we find a single physical state, with $q=\ha(\p^+{-}\p^-)$.
In fact, there is a subtlety in defining a local spectral-flow operator.
Namely, $\sfo(\q,z)$ must depend on an arbitrary reference point~$z_0$,
where all charges are reversed compared to those at~$z$.
Fortunately, $z_0$~dependence drops out if the spectral flows 
of all vertex operators in the correlator sum to zero.
Given that $\partial\sfo(\q)$ is BRST-exact, 
the correlator is even invariant under such a flow,
allowing one to freely change the relative picture numbers 
of its vertex operators~\cite{kl}.
Somewhat surprisingly, the spectral flow angle~$\q$ is not a compact variable.
Instead, $\ico\equiv\sfo(\q{=}1)$ creates a Maxwell instanton at the
location of the vertex operator.
Thus, an overall spectral flow of $\sum_{i=1}^n\q_i=\D c\in Z$
does change the amplitude and connects different instanton sectors.
The above table then implies that the amplitude $A^g_{c_L,c_R}$ carries
a flavor charge of~$q=c_L{+}c_R$.
Moreover, the worrisome $z_0$~dependence of an instanton-changing flow
has a beautiful interpretation~\cite{buck}:
It is nothing but the inherent ambiguity in comparing different topologies in 
the first-quantized approach and must be absorbed in the string coupling~$\l$.
Consequently, $\l$ must also carry the compensating unit of $q$~charge!
The picture numbers, in contrast, are already balanced 
due to the selection rule 
$$
\sum_{i=1}^n\ (\p^+,\p^-)\ =\ (2g{-}2{-}c\ ,\ 2g{-}2{+}c) \quad,
\eqno(7) $$
and any such picture-number assignment to the vertex operators 
leads to the same amplitude.

Finally, we employ the picture-changing operators~$\pco^\pm$
to map the known BRST cohomology for $\p{=}{-}2$ to that in higher pictures.
Regrettably, this map is not one-to-one, because picture-changing cannot
be inverted for the $N{=}2$ string~\cite{bkl,lp}.
Indeed, $\pco^+$ and $\pco^-\ico$ map to the same picture but differ
by one unit in the flavor charge~$q$!
Hence, starting from $\p{=}{-}2$ and $q{=}0$ 
and successively applying $r$~picture-changing operators,
we have $r{+}1$ possibilities,
$$
(\pco^+)^r\;,\quad
(\pco^+)^{r-1}\pco^-\;,\quad
(\pco^+)^{r-2} (\pco^-)^2\;,\quad\ldots\;,\quad
(\pco^-)^r \quad,
\eqno(8) $$
leading to a spectrum of $q{=}0$ states with 
$|\p^+{-}\p^-|\le r$ at $\p=r{-}2$.
Using spectral flow we may move all these states to the same picture
$(\p^+,\p^-)$ and arrive at $r{+}1$ distinctive states
$$
\ket{\,q={-}\p^-{-}1}\;,\quad
\ket{\,q={-}\p^-}\;,\quad\ldots\;,\quad
\ket{\,q=\p^+}\;,\quad
\ket{\,q=\p^+{+}1}
\eqno(9) $$
with a range of $U(1)_f$ charges, as sketched in fig.~1. 
We will argue that all those states are proportional to one another,
with $q$-charged functions of momentum~$k^\m$ as proportionality factors.
For generic momenta then, the states of eq.~(9) either are all BRST-exact
or represent different cohomology classes.
Since in the first case the three-point function vanishes for $g\ge(r{-}2)/4$,
we assume the second variant to hold.
This scheme suggests that the number of physical states depend on the picture
and no physical states exist in subcanonical pictures $\p<{-}2$.

\begin{figure}[ht]
\begin{center}
\leavevmode\epsfxsize=10cm
\epsfbox{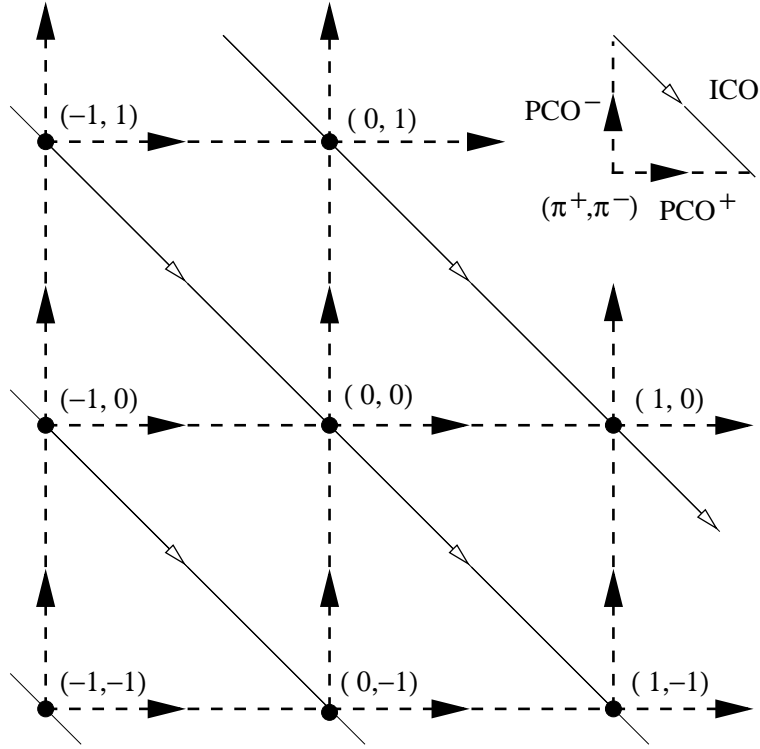}
\caption{Picture-Changing and Spectral Flow}
\end{center}
\end{figure}

It is instructive to look at the simplest example, $r=1$.
We use the following notation.
Complex momenta $k^\m$, complex string coordinates $X^\m$, and 
complex NSR fermions $\j^\m$, with Lorentz index $\m=0,1$,
are expressed in real bispinor notation with respect to the 
$SU(1,1)_f$ and $SU(1,1)_c$ factors of the would-be Lorentz group. 
For instance,
$$
k^{q,m}\ \in\ \{ k^{++}, k^{+-}, k^{-+}, k^{--} \} \qquad{\rm with}\quad
q=\pm\ha \quad{\rm and}\quad m=\pm\ha
\eqno(10) $$
transforms as a $(2,2)$ of $(SU(1,1)_f,SU(1,1)_c)$.
The masslessness condition reads
$$
k^{++}k^{--}\ +\ k^{+-}k^{-+}\ =\ 0 \quad.
\eqno(11) $$
Sometimes we hide the color index~$m$ and simply write~$k^q$.
Denoting the superconformal ghost vacuum at momentum~$k$ by
$\ket{\p^+,\p^-;k}$, one finds 
$$
\begin{array}{c c c c c}
& \raise-1.ex\hbox{${\scriptstyle\pco^-}\nearrow$} & 
k^+{\cdot}\j^-\ket{{-}1,0;k} & & k^-{\wedge}\j^-\ket{{-}1,0;k} \\
\ket{{-}1,{-}1;k} & & \hfill{\scriptstyle(\ico)^{\pm1}} & \cross & \\
& \raise 1.ex\hbox{${\scriptstyle\pco^+}\searrow$} & 
k^-{\cdot}\j^+\ket{0,{-}1;k} & & k^+{\wedge}\j^+\ket{0,{-}1;k} \\
\end{array}
\eqno(12) $$
where we introduced the color contractions 
$$
a \cdot b\ :=\ a^+ b^- + a^- b^+ \quad,\qquad
a \wedge b\ :=\ a^+ b^- - a^- b^+
\eqno(13) $$
among spinors.
Although the two physical states in the lower row of eq.~(12)
differ by one unit in flavor charge~$q$, they are proportional because
$$
\eqalign{
k^+\wedge \j^+\ 
&=\ k^{++}\j^{+-} - k^{+-}\j^{++} \cr
&=\ \frac{k^{++}}{k^{-+}} \left( k^{-+}\j^{+-} + k^{--}\j^{++} \right) \cr
&=\ \frac{k^{++}}{k^{-+}}\ k^-\cdot\j^+ \quad, }
\eqno(14) $$
with the help of eq.~(11).
This proportionality has been exploited in refs.~\cite{bv2,ber}
to derive vanishing theorems for scattering amplitudes.

BRST analysis in the $(0,{-}1)$ picture
also reproduces the lower row of eq.~(12).
The requirement of BRST invariance for physical states
simply restricts the ansatz $\e{\cdot}\j^+\ket{0,{-}1;k}$
by $k^+{\cdot}k^-{=}0$ and $\e{\cdot}k^+{=}0$.
The general solution to the second condition is 
$\e=v^+k^-{\cdot}-v^-k^+{\wedge}$
with free coefficients $v^\pm$, or
$$
\eqalign{
\ket{0,{-}1;{\rm phys}}\ 
&=\ (v^+k^-{\cdot}\j^+ - v^-k^+{\wedge}\j^+) \ket{0,{-}1;k} \cr
&=\ \left( v^+ - \frac{k^{++}}{k^{-+}} v^- \right) 
k^-{\cdot}\j^+ \ket{0,{-}1;k}\quad. }
\eqno(15) $$
More generally, in any picture one finds a $(\p{+}3)$-plet of physical states
which are all proportional to one another but carry different flavor charges.

\section{ Rotating the Complex Structure }

We have made explicit use of a complex structure in spacetime, 
e.g. when formulating the BRST cohomology problem.
As already mentioned, such a choice corresponds to a breaking of
$SU(1,1)_f$ to $U(1)_f$. 
Thus, the two-dimensional moduli space of complex structures
is given by the pseudo-sphere $SU(1,1)_f/U(1)_f$.
In order to reconstruct full $SU(1,1)_f$ symmetry, we must understand
how objects with different $U(1)_f$ charge~$q$ combine to multiplets
under flavor rotations.

Let us parametrize $SU(1,1)_f$ by complex $2\times2$ matrices~\footnote{
The entries $\a$ and $\b$ correspond to the harmonic variables
$u_1$ and $u_2$ of ref.~\cite{bv2,ov2,ber}. 
As they parametrize Lorentz transformations, 
we identify left- and right-moving flavor rotations.}
$$ 
U\ =\ \left(\begin{array}{c c} \a & \b \\ \b^* & \a^* \end{array}\right)
\qquad{\rm with}\quad \a^*\a - \b^*\b = 1 \quad.
\eqno(16) $$
The compact $U(1)_f$ subgroup is given by~$\b=0$.
Decomposing $k^{q,m}$ of eq.~(10) into two flavor spinors~$k^{(m)}$, 
$$
k^{(+)}\ :=\ 
\left(\begin{array}{c} k^{++} \\ k^{-+} \end{array}\right) \qquad\qquad
k^{(-)}\ :=\ 
\left(\begin{array}{c} k^{+-} \\ -k^{--} \end{array}\right) \quad,
\eqno(17) $$
one sees that $k^{(m)}\to U\,k^{(m)}$ under a flavor rotation.
The matter fields $X$ and~$\j$ behave in the same way.
Picture-changing and spectral-flow operators may also be combined
to the flavor doublet 
$$
\pco\ :=\ 
\left(\begin{array}{c}+\pco^-\ico^{+1/2}\\-\pco^+\ico^{-1/2}\end{array}\right) 
\eqno(18) $$
carrying $(\p^+,\p^-)=(\ha,\ha)$.
In the process of computing an amplitude, all world-sheet fields
inside the correlators have to be Wick-contracted, leaving a
function of the external momenta only.
Since those Wick contractions are fully $SO(2,2)$ invariant
and momentum factors always arise from 
$\pa^s\contract{8}{X\ e^{ikX}}$ contractions,
the behavior of the NSR field~$\j$ is irrelevant.
Therefore, the flavor properties of an amplitude may be invoked by
transforming only the string coordinates~$X$ and external momenta~$k_i$
but leaving~$\j$ inert.
In the following we shall consider such restricted flavor rotations.
In particular,
$$
\left(\begin{array}{c} k^+\wedge\j^+ \\ k^-\cdot\j^+\end{array}\right) 
\qquad{\rm and}\qquad
\left(\begin{array}{c} k^+\cdot\j^- \\ k^-\wedge\j^-\end{array}\right) 
\eqno(19) $$
form restricted flavor doublets, which appear in eq.~(12).
In higher pictures, the $\p{+}3$ physical states found above
simply manufacture a spin~$(\p{+}2)/2$ flavor representation.

We shall now reinstall $SO(2,2)$ Lorentz symmetry 
separately for left and right movers.
Ignoring possible left-right coupling due to 
$\pa^s\contract{8}{X\ \bar\pa^t X}$ contractions,
the amplitudes factorize as 
$A^g_{c_L,c_R} = \tilde A^g_{c_L} \tilde A^g_{c_R}$.
The picture-number selection rule~(7) 
and the flavor-charge selection rule $q_{\rm tot}{=}c$ imply that
the chiral $n$-point correlator $\tilde A^g_c$ is built from
$n$~canonical flavor-singlet vertex operators with $\p{=}{-}2$ 
together with $r{=}2s{\equiv}4g{-}4{+}2n$ picture-changing operators 
and $c$ instanton-creation operators.
Since $c$ ranges from ${-}s$ to ${+}s$, 
one has $2s{+}1$ chiral correlators for a given~$g$.
Eq.~(18) then strongly suggests that they all combine
in a spin~$s$ flavor representation,
which obtains from the symmetrized tensor product of $2s$
picture-changing doublets~\cite{bv2}.

The only way to produce flavor singlets is to tensor 
the $(2s{+}1)$-plet $(\tilde A^g_{-s},\ldots,\tilde A^g_{+s})$
with its dual representation, 
which must sit in the coefficient factors 
weighing the different instanton contributions.
In other words, the string coupling constants $\k$ and~$\l$ should
flavor-transform non-trivially.
This goal is achieved most easily if we form flavor singlets
already at the level of vertex operators and/or picture-changing operators. 
Since the antisymmetric product of two flavor doublets yields the singlet,
$$
[ a\,,\,b ] \ \ :=\ \ a^+ b^-\ -\ a^- b^+ \quad,
\eqno(20) $$
it suffices to declare that the coefficients $(v^+,v^-)$ in eq.~(15)
form a flavor doublet~$v$, i.e. $v\to Uv$ under $SU(1,1)_f$.
Eq.~(15) can then be rearranged to
$$
\ket{0,{-}1;{\rm phys}}\ =\ 
\left( [ v\,,\,k^{(+)} ]\,\j^{+-} - [ v\,,\,k^{(-)} ]\,\j^{++} \right) 
\ket{0,{-}1;k} \quad,
\eqno(21) $$
making restricted $SU(1,1)_f$ invariance manifest.
More generally, the chiral $n$-point correlator
$$
\tilde A^g\ :=\ \langle\, V_{\rm can}(k_1) \ldots V_{\rm can}(k_n) \quad
[ v\,,\,\pco ]^{4g-4+2n}\, \rangle
\eqno(22) $$
(suppressing anti-ghost zero mode insertions and modular integrations)
produces the complete bunch of~$\tilde A^g_c$ 
and is manifestly $SO(2,2)$ invariant!

The idea is that the monomials of~$v^\pm$ in eq.~(22) 
(and its right-moving counterpart) are simply provided
by the powers of $\k$ and~$\l$ in eq.~(3), since those always match.
We choose the multiplicities to factorize as
$\ell_{c_L,c_R}{=}\ell_{c_L}\ell_{c_R}$ and identify 
$$
\tilde A^g\ =\ \sqrt{\k}^{2g-2+n}\;\sum_c\ell_c\,\l^{-c}\,\tilde A^g_c\quad.
\eqno(23) $$
This prescription determines the multiplicities as~\cite{bv2,ov2}
$$
\ell_c\ =\ {(4g{-}4{+}2n)! \over (2g{-}2{+}n{-}c)!\,(2g{-}2{+}n{+}c)!} \quad.
\eqno(24) $$
More importantly, we read off
$$
v^\pm\ =\ \k^{1/4}\,\l^{\pm1/2}\ 
=\ e^{\fracmm14\VEV{d}\pm\fracmm{i}2\VEV{a}} \quad,
\eqno(25) $$
$$
\sqrt{\k}\ =\ v^+\,v^- \qquad,\qquad \l\ =\ v^+ / v^- \quad,
\eqno(26) $$
which demonstrates that dilaton and axion fields essentially
encode the length and phase of an $SO(2,2)$ Majorana-Weyl spinor.
For later convenience we form the symmetric square $w\equiv v\times v$,
the $SU(1,1)_f$ vector
$$
\left(\begin{array}{c} w^+ \\ w^0 \\ w^- \end{array}\right)\ 
:=\ \left(\begin{array}{c} 
+v^+v^+/\sqrt{2} \\ -v^+v^- \\ 
-v^-v^-/\sqrt{2} \end{array}\right)\
=\ \sqrt{\k}\,
\left(\begin{array}{c} 
+\l^{+1}/\sqrt{2} \\ -1 \\ 
-\l^{-1}/\sqrt{2} \end{array}\right) \quad,
\eqno(27) $$
which is automatically massless, i.e. $w^0w^0+2w^+w^-=0$.
Under $SU(1,1)_c\otimes SU(1,1)_f$ it transforms as a $(1,3)$,
which may be interpreted as a anti-self-dual two-form in (2,2) spacetime.
Converting to a real basis via $w^\pm=(\pm w^1+iw^2)/\sqrt{2}$ and $w^0=-w^3$,
one arrives at eq.~(5).
Apparently, we may $U(1)_f$ rotate away the vev of the axion, 
i.e. set $\l=1$ by a special coordinate choice.
Then, one boost freedom remains to scale the dilaton vev 
to an arbitrary number. 
Yet, different ``observers'' in general will disagree on the size
of both $\VEV{d}$ and~$\VEV{a}$.

\section{ Invariant Amplitudes and Spacetime Actions}

We can now study the change of scattering amplitudes 
under flavor transformations and assemble them into $SO(2,2)$ invariants.
The essential ingredients are the $SU(1,1)_c$ invariant momentum bilinears
$$
\eqalign{
s_{ij}  \ &:=\ k_i^+\cdot k_j^- + k_i^-\cdot k_j^+ \
=\ k_i^{++}k_j^{--}+k_i^{+-}k_j^{-+}+k_i^{-+}k_j^{+-}+k_i^{--}k_j^{++} \cr
c_{ij}^0\ &:=\ k_i^+\cdot k_j^- - k_i^-\cdot k_j^+ \
=\ k_i^{++}k_j^{--}+k_i^{+-}k_j^{-+}-k_i^{-+}k_j^{+-}-k_i^{--}k_j^{++} \cr 
c_{ij}^+\ &:=\ \sqrt{2}\,k_i^+\wedge k_j^+ \qquad\quad\!\!
=\ \sqrt{2}\,( k_i^{++}k_j^{+-} - k_i^{+-}k_j^{++} ) \cr
c_{ij}^-\ &:=\ \sqrt{2}\,k_i^-\wedge k_j^- \qquad\quad\!\!
=\ \sqrt{2}\,( k_i^{-+}k_j^{--} - k_i^{--}k_j^{-+} ) \quad.\cr }
\eqno(28) $$
Under $SU(1,1)_f$, the Mandelstam variable $s_{ij}$ is a singlet,
while $c_{ij}^\e$ form a triplet, $\e=+,0,-$,
as may be checked explicitely from eqs. (16) and~(17).
Complex conjugation exchanges $c_{ij}^+$ and $c_{ij}^-$ while
$c_{ij}^0$ is purely imaginary.
Any scattering amplitude is expressed in terms of $s_{ij}$ and~$c_{ij}^\e$,
and various identities can be derived on-shell, for
$s_{ii}=0=\sum_i k_i$~\cite{ov1,par}.

Since only the three-point function is non-vanishing, let us be
explicit for its chiral half, $\tilde A^g_c(k_1,k_2,k_3)$.
Massless kinematics dictate that all $s_{ij}{=}0$ and 
$c_{i,i+1}^\e{=}-c_{i+1,i}^\e=:c^\e$.
The first non-trivial identity is quadratic,
$$
c^0 c^0 \ +\ 2\,c^+ c^- \ =\ 0 \quad,
\eqno(29) $$
and expresses the lightlike nature of the flavor vector~$c$.
At tree-level, $c_L$ and $c_R$ range from $-1$ to~$+1$.
Straightforward computation of the 
chiral correlators $\tilde A^0_c$ yields~\cite{lp,kl}
$$
\tilde A^0_0\ =\ -\frac12\,c^0 \qquad,\qquad 
\tilde A^0_{\pm1}\ =\ \mp{\textstyle{1\over\sqrt{2}}}\,c^\pm \quad.
\eqno(30) $$
Using $\ell_{\pm1}{=}1$ and $\ell_0{=}2$, this fits perfectly with
$$
\eqalign{
\tilde A^0\ 
&=\ \langle\,V_{\rm can}(k_1)\,V_{\rm can}(k_2)\,V_{\rm can}(k_3) \quad
[ v\,,\,\pco ]^2\,\rangle \cr
&=\ \sqrt{\k}\,\left(
\l^{-1}\tilde A^0_{+1}\ +\ 2\tilde A^0_0\ +\ \l^{+1}\tilde A^0_{-1}\right)\cr
&=\ \sqrt{\k}\,\left(
-{\textstyle{1\over\sqrt{2}}}\,\l^{-1} c^+\ -\ c^0\ +\
 {\textstyle{1\over\sqrt{2}}}\,\l^{+1} c^- \right) \cr
&=\ w^- c^+\ +\ w^0 c^0\ +\ w^+ c^- \cr }
\eqno(31) $$
which is imaginary and manifestly flavor invariant as a scalar product 
of the two lightlike vectors $c$ and~$w$ (see eq.~(27)).

The full tree-level three-point function for the closed string
is simply the square of the above sum,
$$
\eqalign{
A^0\ =\ (\tilde A^0)^2\ 
&=\ w_-^2c_+^2\ +\ 2w_-w_0c_+c_0\ +\ \frac32 w_0^2c_0^2\ +\
2w_0w_+c_0c_-\ +\ w_+^2c_-^2 \cr
&=\ \k\,\left( \ha\l^{-2}c_+^2\ +\ \sqrt{2}\l^{-1}c_+c_0\ +\ 
\frac32 c_0^2\ -\ \sqrt{2}\l^{+1}c_0c_-\ +\ \ha\l^2c_-^{+2}\right)\quad,\cr }
\eqno(32) $$
where we have lowered the flavor superscripts for convenience.
With our conjectured Lorentz behavior~(5) of the string couplings understood,
it is $SO(2,2)$ invariant.

It is well-known that the zero-instanton tree-level amplitudes,
$$
A^0_{0,0}(k_1,\ldots,k_n)\ =\ \d_{n,2}\ +\ \frac14 c_0^2\,\d_{n,3} \quad,
\eqno(33) $$
are reproduced (at least up to $n{=}6$~\cite{par})
by the Plebanski action~\cite{ple} $S_{0,0}=\int d^4x\,L_{0,0}$, with
$$
\eqalign{
L_{0,0}\ &=\ \ha\pa_+\f\bullet\pa_-\f\ 
+\ \frac23\k\,\f\, \pa_+\pa_-\f\,{\wedge}{\wedge}\,\pa_+\pa_-\f \cr
&=\ \ha(\pa_+^+\f\,\pa_-^-\f + \pa_+^-\f\,\pa_-^+\f)\ +\ \frac43\k\,\f\,
(\pa_+^+\pa_-^+\f\,\pa_+^-\pa_-^-\f-\pa_+^+\pa_-^-\f\,\pa_+^-\pa_-^+\f)
\quad. \cr}
\eqno(34) $$
We have defined $\pa_q^m:=\pa/\pa x^{qm}$
and used the notation of eqs. (10) and~(13).
In the double contractions $\pa\pa\f s_1s_2\pa\pa\f$, 
the first symbol ($s_1$) refers to the pairing
$\contract{8}{\pa\pa\f s_1s_2\pa}\pa\f$
while the second symbol ($s_2$) specifies
$\pa\contract{8}{\pa\f s_1s_2\pa\pa}\f$.
Note that this action yields $-\frac12 c^+c^-$ for the three-point function,
which agrees with eq.~(33) due to eq.~(29).

Under flavor rotation, the kinetic term in eq.~(34) is invariant.
The cubic interaction, however, transforms into a linear combination
of $L^{\rm int}_{c_L,c_R}$, with ${-}1\le c_L,c_R\le{+}1$, 
as noted in ref.~\cite{par}.
In particular, one generates
$$
\eqalign{
L^{\rm int}_{1,1}\ 
&=\ -\frac23\k\,\f\,\pa_+\pa_+\f\,{\wedge}{\wedge}\,\pa_+\pa_+\f \cr
&=\ -\frac43\k\,\f\,
(\pa_+^+\pa_+^+\f\,\pa_+^-\pa_+^-\f-\pa_+^+\pa_+^-\f\,\pa_+^-\pa_+^+\f)\cr
L^{\rm int}_{0,1}\ 
&=\ -\frac23\k\,\f\,\pa_+\pa_+\f\,{\wedge}{\bullet}\,\pa_+\pa_-\f \cr
&=\ -\frac23\k\,\f\,
(\pa_+^+\pa_+^+\f\,\pa_+^-\pa_-^-\f+\pa_+^+\pa_+^-\f\,\pa_+^-\pa_-^+\f 
-\pa_+^-\pa_+^+\f\,\pa_+^+\pa_-^-\f-\pa_+^-\pa_+^-\f\,\pa_+^+\pa_-^+\f)
\;.\cr}
\eqno(35) $$
The other interactions obtain from 
$L^{\rm int}_{c_L,c_R}=L^{\rm int}_{c_R,c_L}=L^{{\rm int}\ *}_{-c_L,-c_R}$
and $L^{\rm int}_{1,-1}=L^{\rm int}_{0,0}$.
It is by now obvious that an $SO(2,2)$ invariant action must combine
all these terms with the same coefficients as in eq.~(32).
Expressed in terms of dilaton and axion fields, we finally arrive at
the extended Plebanski action
$$
\eqalign{
S_{\rm inv}[\f,d,a]\ &=\ \int d^4x\;\biggl[ \ha\,\pa_+\f\bullet\pa_-\f\ 
+\ \frac23\,e^{d}\,\f\,\Bigl(\,
6\,\pa_+\pa_-\f\,{\wedge}{\wedge}\,\pa_+\pa_-\f \cr
&\qquad\qquad
-\ 4\,e^{-ia}\,\pa_+\pa_+\f\,{\wedge}{\bullet}\,\pa_+\pa_-\f\ 
-\ e^{-2ia}\, \pa_+\pa_+\f\,{\wedge}{\wedge}\,\pa_+\pa_+\f \cr
&\qquad\qquad
+\ 4\,e^{+ia}\,\pa_-\pa_+\f\,{\wedge}{\bullet}\,\pa_-\pa_-\f\ 
-\ e^{+2ia}\, \pa_-\pa_-\f\,{\wedge}{\wedge}\,\pa_-\pa_-\f 
\Bigr) \biggr] \quad. \cr }
\eqno(36) $$

Let us finally recast this into a more standard notation.
Introducing dotted spinor indices via
$$
\{ k^{++}, k^{+-}, k^{-+}, k^{--} \}\ \equiv\
\{ k^{+\dt{+}}, k^{+\dt{-}}, k^{-\dt{+}}, -k^{-\dt{-}} \}\
\ni\ k^{\a\dt{\a}}
\eqno(37) $$
(see also eq.~(17)), we may use the fact that the only numerically
invariant $SU(1,1)$ tensors at our disposal are
$\d^\a_\b$ and $\e_{\a\b}$ (and their dotted versions).
The bilinears of eq.~(28) read
$$
\eqalign{
s_{ij}\ &=\ -\e_{\a\b}\,
\e_{\dt{\a}\dt{\b}}\ k_i^{\a\dt{\a}}\,k_j^{\b\dt{\b}} \quad,\qquad 
c^0_{ij}\ =\ -(\d^+_\a\d^-_\b+\d^-_\a\d^+_\b)\,
\e_{\dt{\a}\dt{\b}}\ k_i^{\a\dt{\a}}\,k_j^{\b\dt{\b}} \quad,\cr
c^+_{ij}\ &=\ \sqrt{2}\ \d^+_\a\d^+_\b\,
\e_{\dt{\a}\dt{\b}}\ k_i^{\a\dt{\a}}\,k_j^{\b\dt{\b}} \quad,\qquad
c^-_{ij}\ =\ -\sqrt{2}\ \d^-_\a\d^-_\b\,
\e_{\dt{\a}\dt{\b}}\ k_i^{\a\dt{\a}}\,k_j^{\b\dt{\b}} \quad,\cr}
\eqno(38) $$
so that eq.~(31) becomes
$$
\tilde A^0\ =\ 
\o_{\a\b}(d,a)\ \e_{\dt{\a}\dt{\b}}\ k_2^{\a\dt{\a}}\,k_3^{\b\dt{\b}}\ 
=\ \O^-_{mn}(d,a)\ k_2^m\,k_3^n 
\eqno(39) $$
with $m=(\a\dt{\a})$ and $n=(\b\dt{\b})$ denoting $2{+}2$ dimensional
Lorentz indices.
Here, we defined the symmetric degenerate $SU(1,1)$ tensor
$$
\o(d,a)\ :=\ 
\sqrt{\k}\,\left(\begin{array}{cc}-\l^{-1}&1\\1&-\l\end{array}\right)\ =\
e^{d/2}\,\left(\begin{array}{cc}-e^{-ia}&1\\1&-e^{ia}\end{array}\right)
\eqno(40) $$
and its antisymmetric anti-self-dual $SO(2,2)$ cousin
$\O^-_{mn}=\O^-_{\a\dt{\a},\b\dt{\b}}:= \o_{\a\b}\ \e_{\dt{\a}\dt{\b}}$.
The extended Plebanski action of eq.~(36) then takes the form
$$
S_{\rm inv}[\f,d,a]\ =\ \ha\, \h^{mn}\, \pa_m\f\,\pa_n\f\ -\ 
\frac23\ \O_-^{mn}\O_-^{pq}\,\f\,\pa_m\pa_p\f\,\pa_n\pa_q\f \quad,
\eqno(41) $$
with the metric $\h^{\a\dt{\a},\b\dt{\b}}=-\e^{\a\b}\e^{\dt{\a}\dt{\b}}$.
Of course, it would be very interesting to construct a manifestly 
$SO(2,2)$ invariant action $S[g_{mn},d,a]$ for self-dual gravity
with the full 4d metric~$g_{mn}$ from these data.

\vglue.2in\noindent
We acknowledge fruitful discussions with Sergei Ketov and Andrei Losev.

\vfil

\end{document}